\include{aasjournal.bst}
\include{aastex631.cls}
\documentclass{aastex631}
\newcommand{\bfm}[1]{\mbox{\boldmath{$#1$}}}

\usepackage{subfigure}
\usepackage{appendix}

\shorttitle{AASTeX v6.31 Sample article}
\shortauthors{Jackson et al.}

\graphicspath{{./}{figures/}}

\begin{document}

\title{Active Main-Belt Asteroid (6478) Gault - Constraint on Its Cohesive Strength and the Fate of Ejected Particles in the Solar System}

\author[0000-0002-5318-4291]{Pierce M. Jackson}
\affiliation{Department of Aerospace Engineering\\
Auburn University, 211 Davis Hall\\
Auburn, Alabama 36849, USA}

\author[0000-0002-9840-2416]{Ryota Nakano}
\affiliation{Department of Aerospace Engineering\\
Auburn University, 211 Davis Hall\\
Auburn, Alabama 36849, USA}

\author{Yaeji Kim}
\affiliation{Department of Aerospace Engineering\\
Auburn University, 211 Davis Hall\\
Auburn, Alabama 36849, USA}

\author{Masatoshi Hirabayashi}
\affiliation{Department of Aerospace Engineering\\ 
Auburn University, 211 Davis Hall\\
Auburn, Alabama 36849, USA}
\affiliation{Department of Geosciences\\ 
Auburn University, 2050 Beard Eaves Coliseum\\
Auburn, Alabama 36849, USA}

\begin{abstract}
Active asteroid (6478) Gault sheds mass independent of location along its orbit. Rotational instability is considered to induce the observed activities. If this is the case, because Gault's breakup event hasn't been detected, surface failure is likely, implying its surface materials are constantly ejected while its major body remains intact. Given this scenario, we first constrain Gault’s bulk cohesive strength. We then characterize heliocentric trajectories of ejected particles over thousands of years. The results show that Gault may be sensitive to structural failure at the current spin period ($\sim 2.5\ \textrm{hr}$). Gault's bulk density needs to be below $1.75\ \textrm{g\ {cm}\textsuperscript{-3}}$ in order for particles on the equatorial surface to be shed due to centrifugal forces. In this case, Gault requires cohesive strength of at least $\sim 200\ \textrm{Pa}$ to maintain the structure at the center, whereas the surface strength needs to be less than $\sim 100\ \textrm{Pa}$ to induce mass shedding. This suggests that Gault's structure may consist of a weak surface layer atop a strong core. The trajectories of dust ejected from Gault depend on how efficiently they are accelerated by solar radiation pressure. Escaped particle clouds with sizes of $< \sim 100 \mu$m could collide with Gault after $\sim 700-5,300$ years with speeds of $\sim 0.2$ km/sec. This implies a temporal increase in the impact flux and complex interactions between the ejected particles and their host body.
\end{abstract}

\keywords{minor planets, comets: general - minor planets, asteroids: general - minor planets, asteroids: individual (6478 Gault)}

\section{Introduction} 
\label{sec:intro}

Active asteroids are those having unique activities but are distinguished from comets \citep{jewitt2012active,hsieh2020active}. While there is no distinctive parameter that can perfectly distinguish them from comets, dynamic parameters such as the Tisserand parameter with respect to Jupiter ($T_J$) are good indicators to do so \citep{jewitt2012active,hsieh2020active}. The presence of dust tails is one of the most common activities to be observed among active asteroids. Some have continuous dust tails, which imply continuous mass ejection processes, while others exhibit episodic tails, which are indicative of discontinuous mass ejection processes \citep{jewitt2015episodic,jewitt2019episodically}. Many different processes might contribute to dust tail production. Potential sources to induce mass ejection processes include rotational instability (i.e., P/2013 R3 (Catalina-PANSTARRS) and 311P/PANSTARRS) \citep{jewitt2017anatomy,jewitt2018nucleus}, sublimation of volatiles (i.e., 133P (7968 Elst-Pizarro) and 238P) \citep{hsieh2010return,hsieh2011main}, and thermal fatigue (i.e., Phaethon) \citep{Jewitt2010}. The suggested activities of these asteroids are based on earlier hypotheses, although detailed evidence is still lacking. Asteroid 311P generates episodic dust tails \citep{jewitt2015episodic} possibly because it is part of a binary system and they rob each other of material \citep{hainaut2014continued, hainaut2014p,jewitt2018nucleus}. There are also other processes that could induce such activities \citep[e.g.,][]{jewitt2012active, hsieh2020active}. 

One of these processes is rotational instability. When an asteroid rotates beyond its spin limit, materials experience strong centrifugal forces exceeding gravity, cohesion, and adhesion and thus are ejected from a surface while the interior is structurally intact (surface failure). The internal structure can also be disturbed, causing a large-scale reshaping process (internal failure) (\cite{Hirabayashi2015a, hirabayashi2015failure, sanchez2018rotational}). These processes can induce mass ejection events, but the magnitude of these events may differ from one asteroid to another. Surface failure, in general, causes smaller activities than internal failure. This is because if internal failure occurs, the body could break up completely. A good example is P/2013 R3, which broke up and exhibited multiple components escaping from each other in 2013 \citep{hirabayashi2014constraints, jewitt2017anatomy}. Surface failure can occur even at rotational periods below the spin limit, initiating mass movement on the surface \citep{hirabayashi2020spin}. In this case, materials are not ejected but may move over the surface as seen on asteroid Bennu (\cite{walsh2019craters, jawin2020global, daly2020hemispherical}). While there are many parameters that contribute to structural failure, the cohesive strength is considered to be a primary contributor (\cite{scheeres_cohesion, strength_sanchez_2014}). In continuum mechanics, this usually represents shear strength of zero pressure \citep{Chen1988plasticity}. Recent work has shown that typical cohesive strengths of small bodies ($<1$ km in diameter) are likely around few hundred Pa \citep{hirabayashi2014constraints,nakano2020mass}. Recent investigations of asteroids (101955) Bennu and (162173) Ryugu by OSIRIS-REx \citep{barnouin2019shape} and Hayabusa2 \citep{watanabe2019hayabusa2,roberts2021rotational}, respectively, showed very low cohesive strengths on their surfaces (less than $\sim 10$ Pa), although no upper bounds of the bulk cohesive strength were provided. 

Main belt asteroid (6478) Gault is an active asteroid classified into the Phocaea family \citep{sanchez2019physical}. Gault has a Tisserand parameter with respect to Jupiter of $T_J$ = 3.46, which is among typical asteroids ($T_J >$  3.0) \citep{hsieh2020active}. This asteroid exhibits episodic dust tails \citep[e.g.][]{tonry2018atlas, hui2019new, kleyna2019sporadic, moreno2019dust, ye2019multiple} independent of locations along its orbit \citep{chandler2019six}, implying that the observed activities are not related to volatile sublimation and thermal alteration. Since September 2013, multiple activities have been observed to continue for up to 20 days \citep{jewitt2019episodically, chandler2019six,ye2019multiple}, which likely rules out the possibility that these events are due to independent meteoroid impacts \citep{ishiguro2011observational}. This asteroid is also reported to have a bulk composition similar to an S-type \citep{sanchez2019physical,marsset2019active}. \citet{purdum2021time} recently obtained Gault's low lightcurve amplitudes and inferred that this asteroid might have a top shape (a spheroidal object having an equatorial ridge). However, it is still uncertain whether this asteroid is indeed a top shape due to limited data. \citet{purdum2021time} also conducted spin analysis, but the axis ratios used in their analysis were an assumption based on earlier work \citep{harris2014maximum}. Although still uncertain at present, Gault’s shape will be further constrained by future observations.

Rotational instability could be one of the plausible explanations for Gault's activities \citep{devogele20216478}. The most recent estimation of its spin period is $\sim$2.5 hr \citep{devogele20216478, purdum2021time, luu2021rotational}. This spin condition is close to the spin barrier ($\sim$2.2 hr) that a rubble pile body becomes structurally unstable due to a high centrifugal force \citep{pravec2000fast}. If rotational instability is indeed a cause of the observed mass ejection, surface failure is a likely scenario because there is no evidence of a breakup \citep{sanchez2016disruption}. If this is the case, surface materials are ejected during the activities while the major body remains intact. In fact, the magnitude of observed mass ejection events have been estimated to be $\sim 1\times 10^6$ - $2\times 10^7$ kg, which is negligible compared to the asteroid's total mass, $\sim 10^{13}$ kg, supporting the surface failure scenario \citep{ye2019multiple,moreno2019dust}. Also, for the activities to occur, particles must be ejected at speeds greater than Gault's escape velocity. Thus, the spin period should exceed the spin limit, constraining the geophysical conditions.

In this study, we investigate Gault's structural conditions and surrounding particle environment driven by mass shedding processes by applying the physical properties estimated by observations and assuming a spherical shape with uniform bulk density. First, we analyze the minimum cohesive strength required such that Gault avoids internal failure but experiences surface failure at the current spin period. We then employ a simplified N-body dynamics simulation to investigate if ejected particles come back to Gault on astronomically short time scales ($<$ 6,000 years). By these two independent approaches, we discuss how Gault's mass ejections connect with its structure and surrounding dust environment.

\section{Semi-analytical structural model} \label{sec:2}

We employ a semi-analytical model, developed by \cite{nakano2020mass}, which determines Gault's failure mode at the current spin period $P$. In the following discussions, we assume that Gault is a sphere with a diameter of 2.80 km \citep{devogele20216478}, given this asteroid's observed low amplitudes \citep{purdum2021time} and the current uncertainties of its shape and axis ratios. Unlike a finite element model approach \citep[e.g.,][]{hirabayashi2020spin}, the semi-analytical model does not capture the detailed shape; however, considering the shape to be a sphere is still effective to predict the global cohesive strength distribution of an asteroid without concavity \citep{hirabayashi2015failure}. We also note that the motivation of using a simplified model came from \citet{hirabayashi2020spin}'s observation that the consideration of equatorial ridges do not affect the overall cohesive strength calculation. Nevertheless, in Appendix \ref{App:non_spherical}, we discuss two example cases if Gault were to have an ellipsoidal shape: one with the same axis ratios as (101955) Bennu and one with the same axis ratios as (1580) Betulia \citep{hirabayashi2019rotationally}.

$P$ is set as $2.50 \ \text{hr}$ based on earlier reports suggesting $2.4929 \pm 0.0003 \ \text{hr}$ by \cite{devogele20216478}, $\sim 2.5 \ \text{hr}$ by \cite{purdum2021time}, and $2.55 \pm 0.10 \ \text{hr}$ by \cite{luu2021rotational}. We assume that Gault’s bulk density, $\rho$, is uniform and ranges from $1.50\ \textrm{g\ {cm}\textsuperscript{-3}}$ to $3.00\ \textrm{g\ {cm}\textsuperscript{-3}}$ to cover Gault's potential taxonomic class: S-type \citep{sanchez2019physical, marsset2019active}. We use Poisson’s ratio, $\nu$, and Young’s modulus, $E$, to determine the stress field; these quantities are defined as $0.25$ and $10^7\ \textrm{Pa}$, respectively \citep{holsapple2007spin, hirabayashi2014stress}.

Our approach first computes the stress in Gault with uniform rotation by applying Hooke's law and the traction free boundary condition \citep{dobrovolskis1982internal, holsapple2001equilibrium, hirabayashi2015failure, nakano2020mass}. In this case, the stress field is independent of $E$. The obtained stress field inside the body is then used to determine the structural failure condition, which is characterized using the Drucker-Prager yield criterion \citep{Chen1988plasticity}:

\begin{equation}
    f = \alpha I_1 + \sqrt{J_2} - s \leq 0. \label{Eq:D-P}
\end{equation}

\noindent where $I_1$ and $J_2$ are the stress invariants:

\begin{equation}
    I_1 = \sigma_1 + \sigma_2 + \sigma_3,
\end{equation}

\begin{equation}
    J_2 = \frac{1}{6} \{(\sigma_1 - \sigma_2)^2 + (\sigma_2 - \sigma_3)^2 + (\sigma_3 - \sigma_1)^2\},
\end{equation}

\noindent and $\sigma_i$ ($i = 1, 2, 3$) is the principal stress component. $\alpha$ and $s$ are material constants and given as \citep{Chen1988plasticity}:

\begin{equation}
    \alpha = \frac{2 \sin{\phi}}{\sqrt{3} (3-\sin{\phi)}},
\end{equation}

\begin{equation}
    s = \frac{6Y \cos{\phi}}{\sqrt{3} (3-\sin{\phi)}},
\end{equation}

\noindent where $Y$ denotes the cohesive strength, and $\phi$ denotes the friction angle and is fixed at $35\ \textrm{deg}$ in this study \citep{Lambe1969}. 

The critical cohesive strength is given by setting Equation (\ref{Eq:D-P}) equal to zero and solving for $Y$. Here, it is denoted as $Y^*$ to be distinguished from the actual cohesive strength $Y$ that the body actually has. We obtain the following expression for $Y^*$:

\begin{equation}
    Y^* = \frac{\sqrt{3} (3 - \sin{\phi})}{6\cos{\phi}}\left(\alpha I_1 + \sqrt{J_2} \right). \label{Eq:Cohesive}
\end{equation}

\noindent If Equation (\ref{Eq:Cohesive}) becomes negative, we set $Y^*$ as $0 \ \textrm{Pa}$, which means that no cohesive strength is required for an element to avoid failure and inelastic deformation. On the other hand, if $Y^* > 0 \ \textrm{Pa}$, an element requires a cohesive strength $Y$ larger than $Y^*$ to avoid structural failure. The parameters used for this model are provided in Table \ref{table:1}. 
\section{Dynamics of particles ejected from Gault}
\label{sec:3}
We quantify the motion of particles ejected from Gault for a short time period. Instead of using well-established simulation tool such as SWIFT \citep{1991AJ....102.1528W,1994Icar..108...18L,1991CeMDA..52..221G}, or MERCURY \citep{1999MNRAS.304..793C}, we develop a simplified N-body simulation tool to analyze particle dynamics. As our scope is a short-term evolution of ejected particles, we do not anticipate complex dynamic effects causing particles' chaotic behaviors. The dust dynamics model considered is similar to the one used by \citet{POPPE2016369} and accounts for the gravitational accelerations on the particles by the Sun and 14 planetary bodies (including all planets, Earth's moon, Pluto and Charon, Ceres, Vesta, and Gault), as well as the non-gravitational accelerations by solar radiation pressure (SRP) and Poynting-Robertson drag (PRD). The orbital motion of a particle, which is here denoted using index $i$, is given by 
\begin{eqnarray}
\ddot {\bfm r}_i = {\bfm a}^g_i + {\bfm a}^{pert}_i \label{Eq:EOM}
\end{eqnarray}
where $\bfm r_i$ is the heliocentric position vector, ${\bfm a}^g_i$ is the gravitational acceleration, and ${\bfm a}^{pert}_i$ is the non-gravitational acceleration. 

For ${\bfm a}^g_i$, we consider particle $i$ to be influenced by the bodies mentioned above, which are defined using index $j$, and written as
\begin{eqnarray}
    {\bfm a}_i^g = - \sum_{j = 1}^{j_{max}} GM_j \frac{{\bfm r}_i - {\bfm r}_j}{\vert {\bfm r}_i - {\bfm r}_j \vert^3}
    \label{Eq:Gravity}
\end{eqnarray}
where $G$ is the gravitational constant, and $M_j$ is the mass of body $j$. In our model, $j_{max} = 15$. On the other hand, ${\bfm a}^{pert}_i$ is given as \citep{rad_forces}
\begin{equation}
    {\bfm a}^{pert}_i = \frac{Q_{pr} L A_{i}}{m_i c} \left[ \left(1 - \frac{\dot r_i}{c} \right)\bfm {\hat S} - \frac{\dot {\bfm r_i}}{c} \right]
    \label{eq:SRPPRD}
\end{equation}
where $m_i$ is the particle mass, $L$ is the energy flux density of the radiation field (i.e. irradiance from the Sun), $\bfm {\hat S}$ is the unit radial vector from the Sun (outward positive), $A_{i}$ is the cross-sectional area of the particle, $c$ is the speed of light, $Q_{pr}$ is the radiation pressure efficiency of the particle (via the Mie theory), and $\dot r_i$ is the radial component of the velocity vector (i.e., $\dot r_i = \dot {\bfm r_i} \cdot \hat {\bfm S}$). The radial term is SRP, and the velocity dependent term is PRD. Because the dust's light absorption and scattering properties are not well constrained, we use three different $Q_{pr}$ values: 0, 1, and 2. The zero value means SRP and PRD are not active - only gravity influences particle dynamics; the unity value indicates a totally absorbing particle and is widely used in the literature \citep{moreno2019dust, POPPE2016369, rad_forces}; and the value of 2 may be too high but nevertheless is considered to see $Q_{pr}$'s effects on dynamics. The radiation pressure efficiency dependence on the size of silicate particles is illustrated in Figure 3 in \citet{trojandust}.

We use a Runge-Kutta 4th-order integrator to solve Equation (\ref{Eq:EOM}), with a time step, $h$, of 432 s which is obtained by dividing the number of seconds in a day by 200 ($0.005 \cdot 86400$ s). We choose this value because of its balance of accuracy and efficiency. We compared simulation results using this 432 s time step against smaller time steps such as 86 s and 43 s. The 432 s time step produced trivial differences in particle trajectories compared to the smaller time steps. Particle-particle interactions (gravitational, electrostatic, frictional forces and collisions) are also not considered - although such interactions may change the behavior of individual particles - but may not affect the statistical locations of ejecta clouds significantly over these short timescales \citep{GRUN1985244, Borin_asteroidal_flux}. We confirmed that the quasi-total energy of a test particle under ${\bfm a}^g_i$ without ${\bfm a}^{pert}_i$ (Equation (1.29) in \citet{Morbidelli2002ModernCM}) was constant with negligible fluctuation.

We define the following simulation settings. First, we obtain the initial ephemerides using the JPL Horizons web-interface tool (\url{https://ssd.jpl.nasa.gov/horizons.cgi}). We then generate dust ejection conditions on September 28, 2013, June 10, 2016, November 12, 2017 \citep{chandler2019six}, October 18, 2018, December 24, 2018 \citep{ye2019multiple, moreno2019dust, jewitt2019episodically}, and February 10, 2019 \citep{jewitt2019episodically}. For a given value of $Q_{pr}$ and a given epoch, we simulate the motion of 1,080 particles departed from Gault for a maximum of 6,000 years to model a mass ejection event.If a cloud of particles returned to Gault before the 6,000 year limit, we terminate the simulation to save time and computational resources. While the number of test particles is lower than usual numerical studies for the orbital evolution of particles, this low resolution simulation still meets our scope that gives general trends within a short period. All particles share a common initial ejection speed of $v = 0.15$ m s$^{-1}$ which was determined by \citet{jewitt2019episodically}. 

Within the 1,080 particles, however, we vary the ejection directions and particle sizes. We divide this total particle number into 10 separate simulations, each of which has 108 particles. In each case, particles are ejected radially out from Gault's surface; 36 at the equator with a 10$^\circ$ separation interval and similarly, 36 from both $\pm 45^\circ$ latitude. The major reason for this setting is the lack of the known spin orientation of Gault. Their sizes are all generated using the broken power law distribution observed by \citet{jewitt2019episodically}. Optical observations implied that the particles in the dust tails from Gault likely have radii, $a_p$, between 1 $\mu$m and 1 mm \citep{jewitt2019episodically, moreno2019dust, ye2019multiple}. All simulations produce an average particle radius of $\bar a_p \sim 180$ $\mu$m which is consistent with \citet{jewitt2019episodically}. Once each individual particle size is set, the mass is computed given a constant density of $\rho_p = 2.5 \textrm{g\ {cm}\textsuperscript{-3}}$ and a spherical shape. Simulation and particle material parameters are given in Table \ref{table:1}.

\section{Results}
\subsection{Structural condition}

Our structural model indicates that Gault requires cohesion, but its failure mode changes depending on the level of cohesive strength and the bulk density. Given Gault's $2.5\ \textrm{hr}$ spin period and bulk density ranging between $1.50\ \textrm{g\ {cm}\textsuperscript{-3}}$ and $3.00\ \textrm{g\ {cm}\textsuperscript{-3}}$, the structural model predicts the critical cohesive strength $Y^* > 0$ within the body. Figure \ref{Fig: structure} shows the distribution of $Y^*$ on the $x$-$z$ plane (the cross section along the spin axis) for four different $\rho$ cases. For the case of $\rho = 1.50\ \textrm{g\ {cm}\textsuperscript{-3}}$ (Figure \ref{Fig: structure}(a)), most of the body has $Y^* > 0$. The maximum $Y^*$ is found to be $\sim 225\ \textrm{Pa}$ at the center of the body, meaning that the central region requires cohesive strength higher than $\sim 225\ \textrm{Pa}$ to remain intact structurally and thus is more sensitive to failure than the surface. A similar trend can be observed for the case of $\rho = 1.75\ \textrm{g\ {cm}\textsuperscript{-3}}$ (Figure \ref{Fig: structure}(b)). The majority of the regions still have high $Y^*$, with the maximum of $\sim 200\ \textrm{Pa}$ at the core. In contrast, we find a different trend for the cases of $\rho = 2.50\ \textrm{g\ {cm}\textsuperscript{-3}}$ and $\rho = 3.00\ \textrm{g\ {cm}\textsuperscript{-3}}$ (Figure \ref{Fig: structure}(c) and (d), respectively). $Y^*$ at the surface is higher than that of the interior. The equatorial surface region exhibits the maximum $Y^*$ for both cases ($\sim 90\ \textrm{Pa}$) and is the most sensitive to structural failure. The interior does require some cohesive strength for the case of $\rho = 2.50\ \textrm{g\ {cm}\textsuperscript{-3}}$ but does not require any cohesive strength for the case of $\rho = 3.00\ \textrm{g\ {cm}\textsuperscript{-3}}$. Given that Gault's activities do not appear to be catastrophic, the present results imply that the interior is probably still intact. This result suggests that Gault's structure may consist of a structurally strong core and a weak surface layer on top of it if this asteroid has a typical bulk density of S-type asteroids, $< 2.5$ g/cm$^3$ \citep{fujiwara2006rubble, yeomans2000radio}. In this case, surface shedding may occur at middle latitudes if surface cohesion is low, while it may occur at the equator if the cohesion is high. This interpretation is consistent with \cite{sanchez2020cohesive}. 

To constrain the condition of mass ejection, we also calculate the critical spin period $P_c$ at which a small particle on the equatorial surface gains the centrifugal acceleration larger than the gravitational acceleration and is lifted off from the surface. For a spherical gravitating body, the critical spin period is given by $P_c = 3.3\ \rho^{-0.5} \ \textrm{hr}$, where $\rho$ is expressed in $\textrm{g\ {cm}\textsuperscript{-3}}$ \citep{pravec2000fast}. Figure \ref{Fig: P_c} shows $P_c$ as a function of $\rho$ (red line) and the spin period at which the central point structurally fails (purple line). If $\rho \le 1.75\ \textrm{g\ {cm}\textsuperscript{-3}}$, particles are shed due to high centrifugal force. This value is similar to the bulk density of the S-type asteroid (25143) Itokawa ($1.9\ \textrm{g\ {cm}\textsuperscript{-3}}$) \citep{fujiwara2006rubble}. Given no evidence of catastrophic disruption, the central region should be supported by a cohesive strength higher than $\sim 200\ \textrm{Pa}$, which is consistent with that of observed rubble pile asteroids and comets with a few km in diameter \citep[e.g.,][]{hirabayashi2014constraints, hirabayashi2020spin}. At the current spin period, however, the surface layer should be unstable and induce mass shedding, implying that materials there may be less cohesive than those in the interior (less than $100\ \textrm{Pa}$ based on Figure \ref{Fig: structure}(b)). We note that our simulations considering different elongated shapes predict the increase of $Y^\ast$ up to $\sim 40$\%, leading to an uncertainty of $\sim 100$ Pa (Figures \ref{fig5:Supplemental1} and \ref{fig6:Supplemental2}, although the trend does not change). We conclude that our investigation with the spherical case is meaningful, given the limited information on Gault’s shape and material properties.

\subsection{Dust evolution}
Our numerical simulations show the orbital evolution of particles ejected from Gault with different values of $Q_{pr}$ (Figure \ref{Fig3: particle_dist}). When SRP and PRD do not perturb the ejecta ($Q_{pr} = 0$), they remain close ($< 0.5$ AU) to Gault throughout the entire 6,000 year timescale (Figures \ref{Fig3: particle_dist}(a) and \ref{Fig4: evolution}(c)). Because of their very slow ejection velocities and the absence of perturbations, they remain on trajectories very similar to that of Gault. 

For the $Q_{pr} = 1$ case for the December 2018 event, the mean distance gradually grows and reaches a maximum of 4.3 AU after about 730 years then begins to fall and level out (Figure \ref{Fig3: particle_dist}(b)). The maximum distance reaches a constant of 4.6 AU after 570 years. The minimum curve linearly grows over approximately 920 years and then sharply decreases until it returns to near-zero at the critical time ($t_c$) of 1,160 years. These particles are all quickly blown `behind' Gault (whereas particles remain both in front and behind Gault for the $Q_{pr} = 0$ case). As they distribute over the orbit, the entire cloud slowly migrates retrograde (relative to Gault). As the cloud migrates, the space between particles increases as well until a major arc of the orbit is populated with particles. However, because the leading face (initially the far side of the particle cloud) migrates retrograde quicker than the trailing face (initially the side of the cloud nearest to Gault), there reaches a point where both faces are equidistant from Gault; i.e. creating a major arc along the orbit which is populated with particles, and the remaining minor arc devoid of particles with Gault located at its midpoint. This is represented by the local maximum obtained by the minimum distance (green) in Figure \ref{Fig3: particle_dist}(b) and (c). After this point, Gault begins to approach the leading face of the cloud because of the retrograde motion of the cloud (relative to Gault) and prograde motion of Gault (relative to the cloud). The distance between the leading face of the cloud and Gault eventually reaches a minimum at some critical time $t_c$. Beyond it, the trajectories of the dust and Gault  overlap, and the probability of collision increases (Figure \ref{Fig4: evolution}d). The $Q_{pr} = 2$ case produced similar trends (Figure \ref{Fig3: particle_dist}(c)), but the dust becomes distributed along the orbit at a shorter timescale. The mean distance reaches 4.2 AU after 330 years and begins to oscillate about 2.3 AU, which is approximately the semi-major axis of Gault. The maximum distance reaches a constant value of 4.6 AU after 270 years. The particles should come back to Gault $\sim 550$ years after the ejection. We include an online animated figure in Figure \ref{fig7:animation} of the particle cloud orbital evolution for the December 2018, $Q_{pr} = 2$ case to assist in understanding this description of the particle cloud motion (Appendix \ref{App:Particle_Ejection}).

These times not only depend on $Q_{pr}$, but also the epoch of ejection. This is because Gault's elliptical orbit gives particles different heliocentric velocities depending on where along the orbit they are ejected. Critical times for the $Q_{pr} = 1$ case of the other epochs tested are as follows: September 2018, $\sim 1,000$ years; June 2016, $\sim 700$ years; November 2017 \& October 2018, $\sim 1,300$ years; and February 2019, $\sim 1,100$ years. The relationship that the $Q_{pr} = 2$ case takes roughly half the time of the $Q_{pr} = 1$ case is valid for all epochs we tested.

To summarize, the results for the $Q_{pr} = 1$ and $Q_{pr} = 2$ cases show all particles being blown away from the Sun soon after ejection (that is, behind Gault), regardless of ejection direction and epoch, and later become widely distributed around Gault's orbit. This happens approximately after 1,200 years (Figure \ref{Fig3: particle_dist}(b)) and 550 years (Figure \ref{Fig3: particle_dist}(c)), respectively, after ejection for the December 2018 event. At this point, the probability of particles returning to and colliding with Gault increases; however, the frequency of collisions may depend on the flux of ejected particles. To infer the encounter speeds of particles with Gault, we track particles on return-trajectories (i.e. after they reach the opposite side of Gault in orbit) that approach within a relative distance of $0.02$ AU. The obtained encounter speeds are up to $0.2$ km/s. However, when these particles are far away from Gault ($> 4$ AU), they can reach speeds up to 8 km/s, regardless of ejection epoch.
\section{Discussion and conclusions}
Assuming that the observed activities result from rotational instability and that Gault has a bulk density similar to that of S-type asteroids, we propose that Gault's structure consists of a weak surface layer on top of a strong core. The structural analysis showed that given the $2.5\ \textrm{hr}$ spin period, Gault would require cohesive strength to avoid structural failure in its interior if its bulk density is less than $\sim 2.5\ \textrm{g\ {cm}\textsuperscript{-3}}$. Also, as this asteroid experienced activities without clear evidence of breakup processes, its structural condition is likely surface failure without internal failure (Figure \ref{Fig4: evolution}(a)). Otherwise, the body cannot keep its major structure intact due to the tensile stress condition (Figure \ref{Fig4: evolution}(b)). If this is the case, a bulk density of $1.75\ \textrm{g\ {cm}\textsuperscript{-3}}$ should be an upper bound of Gault's bulk density, so that particles at least in the equatorial regions can be shed due to centrifugal forces. For a bulk density of less than $1.75\ \textrm{g\ {cm}\textsuperscript{-3}}$, the cohesive strength for the surface layers may be less than $\sim 100$ Pa. On the other hand, the interior would require cohesive strength at least twice as strong as the surface layers', $\sim 200$ Pa, to avoid internal failure. Again, our semi-analytical model computes an upper bound for the surface cohesive strength and a lower bound for the interior cohesive strength. We expect that the surface cohesive strength of Gault would actually be smaller, similar to the case of Bennu and Ryugu, where very low cohesive strengths were inferred by OSIRIS-REx and Hayabusa2, respectively. Thus, the actual ratio of the internal cohesive strength to the surface cohesive strength would be higher than $\sim$ 2, as reported by earlier works \citep[e.g.,][]{sanchez2018rotational, Hirabayashi2015a}.

This two-layer structure allows Gault to keep itself from a catastrophic breakup but still experience mass shedding. The interior should remain intact, given its high cohesive strength, so that Gault can survive under the current $2.5\ \textrm{hr}$ spin period. On the other hand, the upper surface layers are structurally weak and fail relatively easily, as denoted by their low cohesive strengths. Surface mass shedding may also be triggered by small-scale events. For example, micrometeoroid impacts, thermal fatigue \citep{granvik2016super}, or tidal effects \citep{kim2021surface} causing small boulders to roll over the surface and kick up mass as it migrates. For the micrometeoroid impact case, further discussions are given below. Once such small perturbation induces mass movement, mobilized particles are shed from the surface due to high centrifugal forces. Because such a failure mode only occurs at a local scale, the total mass ejection may be limited, leading to a similar spin state. Therefore, Gault continues to keep its sensitivity to rotational instability while new perturbations may induce more mass movement, leading to mass ejection (i.e., episodic activities).

Given the lack of information regarding Gault's internal and surface structures, as well as its seemingly random events, it is difficult to quantitatively predict Gault's future activity. Thus, our structural model can only provide qualitative possibilities of Gault’s future activity. If Gault is, in fact, experiencing rotational instability, and has a strong core and weak surface, there are a few possible future activities. First, Gault could continue to experience low-level mass loss events; however, we cannot predict when these events will occur or how large they will be given currently available information. Second, Gault’s activity may become catastrophic if the interior fails. If the spin rate continually increases, the interior will eventually reach a point where the cohesive strength can no longer counteract tensile and shear stresses by fast rotation, leading to a breakup. If YORP is a major contributor, the timescale of such an event may be long, and so it may not happen in the near future. On the other hand, some impulsive inputs such as a large meteorite impact may trigger this type of activity in a near term.

The dynamical analysis predicted that, given a 0.15 m s$^{-1}$ ejection speed relative to Gault due to rotational instability, small particles with sizes of 10s to 100s of $\mu$m being perturbed by SRP and PRD depend on $Q_{pr}$ and the epoch of ejection to spread over regions surrounding Gault’s orbit. For example, dust clouds with $Q_{pr} = 1$ would come back to Gault between 700 - 1,300 years for the recent events (Figure \ref{Fig4: evolution}(d)). After these critical times, there is a higher probability that Gault encounters ejected particles. According to calculations of $Q_{pr}$ for silicate dust by \citet{trojandust} and the particle size distribution observed by \citet{jewitt2019episodically}, the particles ejected from Gault may have $Q_{pr}$ values between 0.8 and 0.4 which would correspond to critical times for the dust cloud to return to Gault (depending on where along the orbit they are ejected) between $\sim 1,300 - 5,300$ years, respectively. On the other hand, particles outside of the size range influenced by SRP and PRD do not easily distribute themselves throughout Gault's orbit and are not likely to return within such a short period (Figure \ref{Fig4: evolution}(c)). 

The enhanced impact flux by ejected particles may persist on short time scales. Particles most likely to impact Gault have intersection speeds below 0.2 km s$^{-1}$, regardless of when they are ejected. This encounter speed of a single particle may not be high enough to induce mass movement on the surface. Given the momentum conservation law, when a dust particle having 1 mm diameter with a bulk density of 2.5 g cm$^{-3}$ hits a particle on Gault having 1 cm diameter with the same bulk density, the resulting speed of the 1 cm diameter particle may be 20 cm s$^{-1}$. Similarly, a 5 cm diameter particle may have a speed of 0.15 cm s$^{-1}$ by such an event. However, if clouds composed of numerous particles bombard Gault, a small amount of dust or boulders on the surface may be mobilized. This could result in a cascade effect inducing greater mass movement that may cause mass shedding. Such enhanced mobilization is observed on asteroids Ryugu and Bennu \citep{sugita2019geomorphology, walsh2019craters}, as well as by experimental tests \citep{murdoch}. If this is the case, active asteroids with frequent activities may be exposed to interplanetary dust particles and micrometeoroid streams originating from themselves (typically called sesquinary events \citep{ZAHNLE2008660}). However, given our limited assessment, we cannot constrain how such returned small particles can influence Gault’s surface condition. Larger boulders should also be ejected from Gault. They are less influenced by SRP and PRD and would exhibit motion similar to the simulated $Q_{pr} = 0$ case, but could eventually come back after longer times than considered here ($> 10,000$ yrs). Because these larger boulders are less infulenced by SRP and PRD, they will not experience the same accelerations due to perturbations and would likely have encounter speeds much lower than that of the small particles discussed above. How such encounters influence Gault's surface condition is unknown. Finally, this may not be the case for much longer time frames when both Gault and ejected particles dynamically evolve due to complex gravitational interactions with larger bodies and non-gravitational effects. 

We finally note the limitation of our models. First, the structural model assumed a uniform density distribution to determine the cohesive strength distribution; however, a different density level changes the stress level, leading to different failure conditions \citep{hirabayashi2014structural, sanchez2018rotational}. Cohesive strength may change due to porosity and thus the density distribution \citep{britt2003asteroid}. We anticipate that a surface layer may have higher porosity, as is observed on Bennu \citep{barnouin2019shape}. Note that this asteroid's interior is also less dense, which possibly resulted from inelastic deformation \citep{scheeres2020heterogeneous}. The dynamical model assumed the ejected particles to be spherical; however, in reality, this is not the case, and their irregular shapes might induce complex non-gravitational forces \citep{Tsuchiyama1125}. Furthermore, in a given simulation, we used the same values of $Q_{pr}$, density, and ejection velocity, although these parameters influence the orbital evolution of the particles. Finally, due to computational constraints, we only consider 1,080 particles per case to focus on the overall trends, but the number is negligible when compared to the observed activities. However, because our results for all $Q_{pr} = 0.4$, 0.8, 1 and 2 cases are: (a) all consistent with one another in regards to particle motion, (b) all indicate that a cloud of particles is capable of returning to Gault after a few hundreds to a few thousands of years, and (c) do not display any anomalous behavior, it is reasonable to consider that the number of particles used in our simulations produce meaningful results. These limitations and settings likely oversimplify the present discussions, and modeling improvement will shed light on this problem further. 

\section{Acknowledgments}
This work is performed under support by NASA/SSW (NNH17ZDA001N/80NSSC19K0548).

\begin{deluxetable}{cCCc}[h!]
\label{table:1}
\tablenum{1}
\tablecaption{Parameter settings for structural computations (top) and particle ejection simulations (bottom)}
\tablewidth{0pt}
\tablehead{\colhead{Parameter} & \colhead{Symbol} & \colhead{Value} & \colhead{Units}}
\startdata
Gault Radius            & R        & 1.40                   & km\\
Current rotation period & P        & 2.50                   & hr\\
Bulk density            & \rho     & 1.50, 1.75, 2.50, 3.00 & g $\cdot$ cm\textsuperscript{-3}\\
Poisson's ratio         & \nu      & 0.25                   & -\\
Elastic modulus         & E        & 10^{7}                 & Pa\\
Friction angle          & \phi     & 35                     & deg\\
\hline
Gravitational constant  & G        & 6.6738 \times 10^{-11} & m\textsuperscript{3} $\cdot$ kg\textsuperscript{-1} $\cdot$ s\textsuperscript{-2}\\
Ejection velocity       & v        & 0.15                   & m $\cdot$ s\textsuperscript{-1}\\
Radii                   & a_p      & 1-1000                 & $\mu$m\\
Average particle radius & \bar a_p & \sim 180               & $\mu$m\\
Density                 & \rho_p.  & 2.50                   & g $\cdot$ cm\textsuperscript{-3}\\
Integration time step   & h        & 432                    & s\\
Solar radiation pressure efficiency & Q_{pr} & 0, 1, 2      & -\\
\enddata
\tablecomments{$E$ does not influence the stress field calculation \citep{love2013treatise}, and the variations in $\nu$ and $\phi$ do not affect our results for geological materials significantly \citep{Lambe1969, hirabayashi2019rotationally}; Ejection velocities and particle radii interpreted from \cite{jewitt2019episodically, luu2021rotational, moreno2019dust, ye2019multiple, devogele20216478}. $Q_{pr}$ values for silicate dust interpreted from Fig. 3 of \citet{trojandust}.}
\end{deluxetable}

\begin{figure}[h!]
    \centering
    \includegraphics[width=0.7\linewidth]{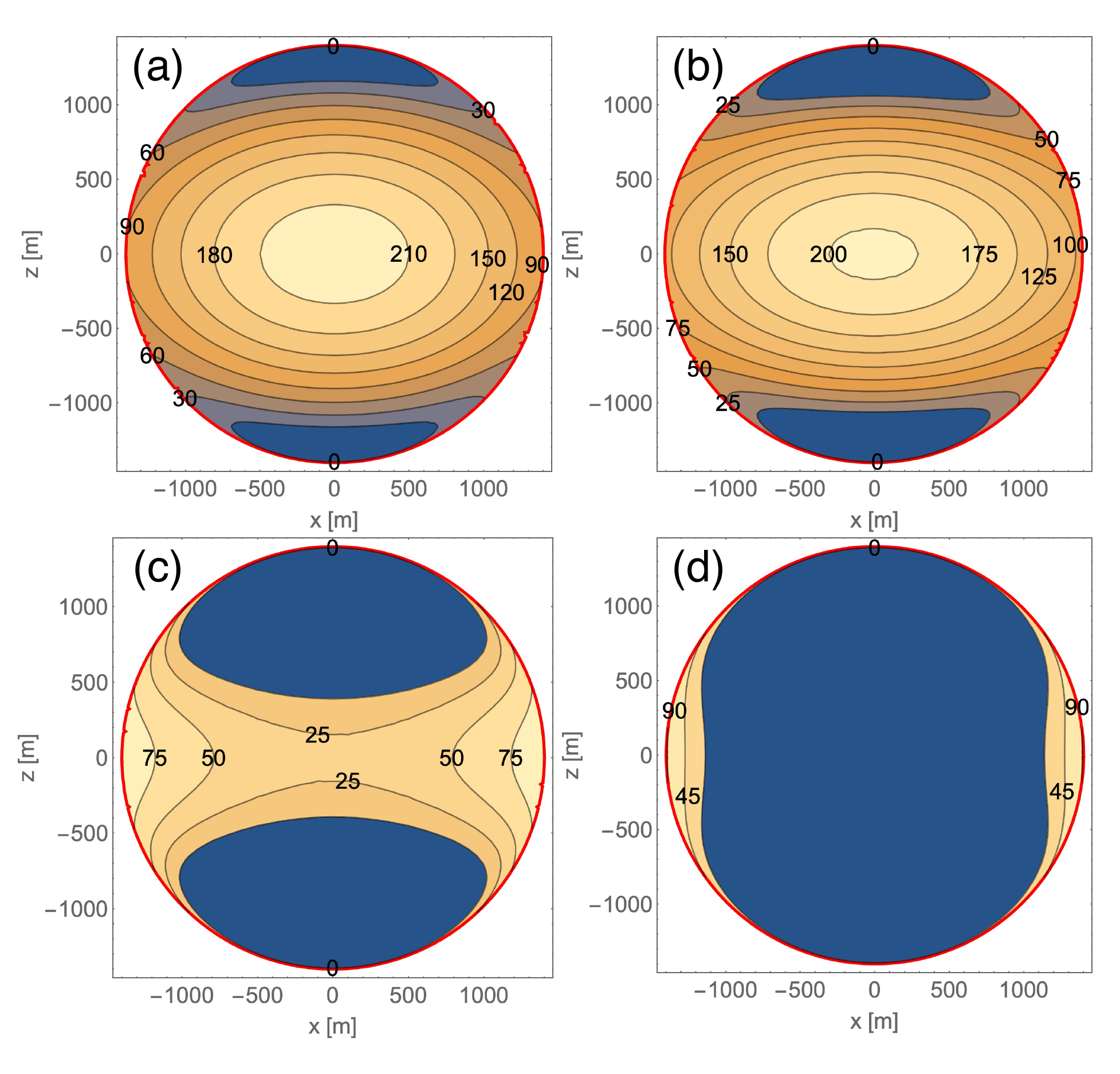}
    \caption{Distribution of $Y^*$ at the $x$-$z$ plane (the cross section along the spin axis). The spin period is $P = 2.5$ hr. Panels a, b, c, and d describe the bulk densities of $1.50$, $1.75$, $2.50$, and $3.00\ \textrm{g\ {cm}\textsuperscript{-3}}$, respectively.}
    \label{Fig: structure}
\end{figure}

\begin{figure}[h!]
    \centering
    \includegraphics[width=0.6\linewidth]{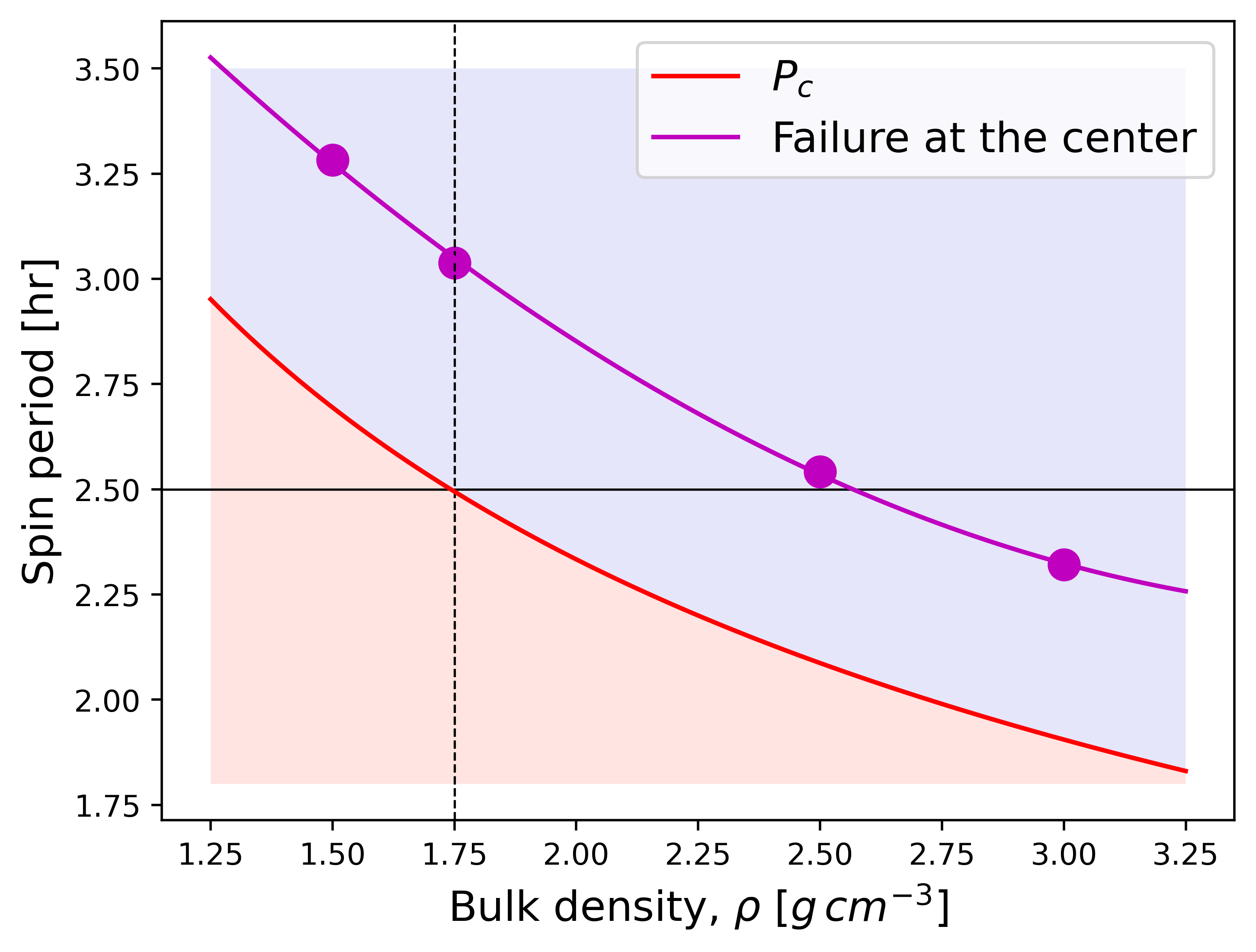}
    \caption{Structural conditions as functions of spin period and bulk density. The red line is the critical spin period $P_c$ as a function of the bulk density. The purple line is the spin period at which the central region structurally fails as a function of the bulk density, which is obtained by polynomial interpolation of the structural model results (purple dots). The gravitational acceleration dominant regime is indicated by the purple shaded region. The centrifugal acceleration dominant regime is indicated by the red shaded region.}
    \label{Fig: P_c}
\end{figure}

\begin{figure}[h!]
    \centering
    \includegraphics[width=0.75\linewidth]{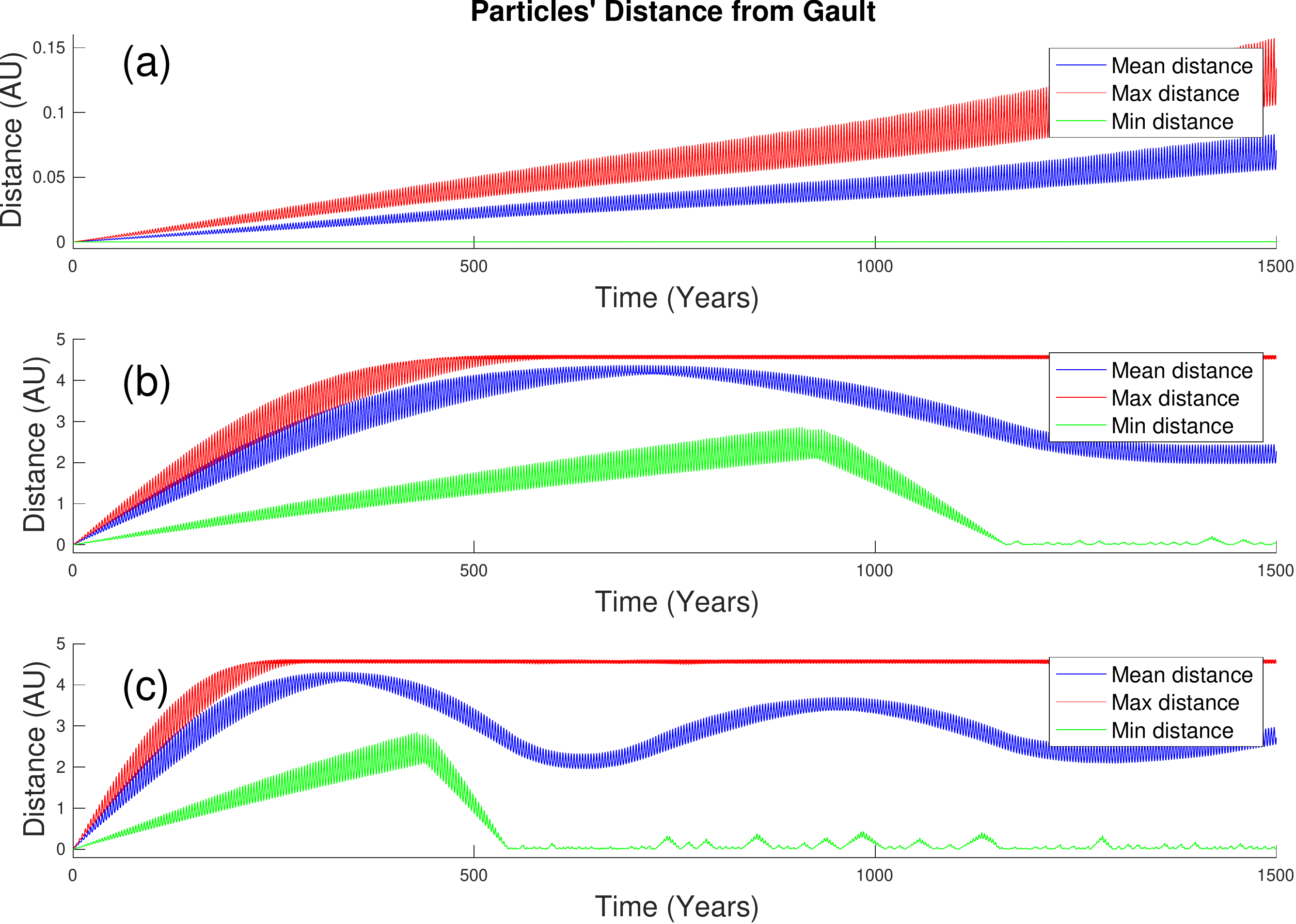}
    \caption{Average particle distance (blue) from Gault of simulated particles ejected during the December 24, 2018 event after 1,500 years for cases: (a) $Q_{pr} = 0$, (b) $Q_{pr} = 1$, and (c) $Q_{pr} = 2$. Red indicates the maximum distance obtained by any particle at that time. Green indicates the minimum distance obtained by any particle at that time. Cases (b) and (c) have critical times ($t_c$) of 1,160 years and 542 years respectively. This is where the minimum distances (green curve) return to near-zero and indicates the amount of time it takes for ejected particles to migrate along Gault's orbit and potentially return to or impact Gault. None of these minimum values reach 0, but have averages of $5.7 \times 10^{-5}$ AU, 0.06 AU (beyond critical time of 1,160 years), and 0.08 AU (beyond critical time of 542 years) for each respective case.}
    \label{Fig3: particle_dist}
\end{figure}

\begin{figure}[h!]
    \centering
    \includegraphics[width=1.0\linewidth]{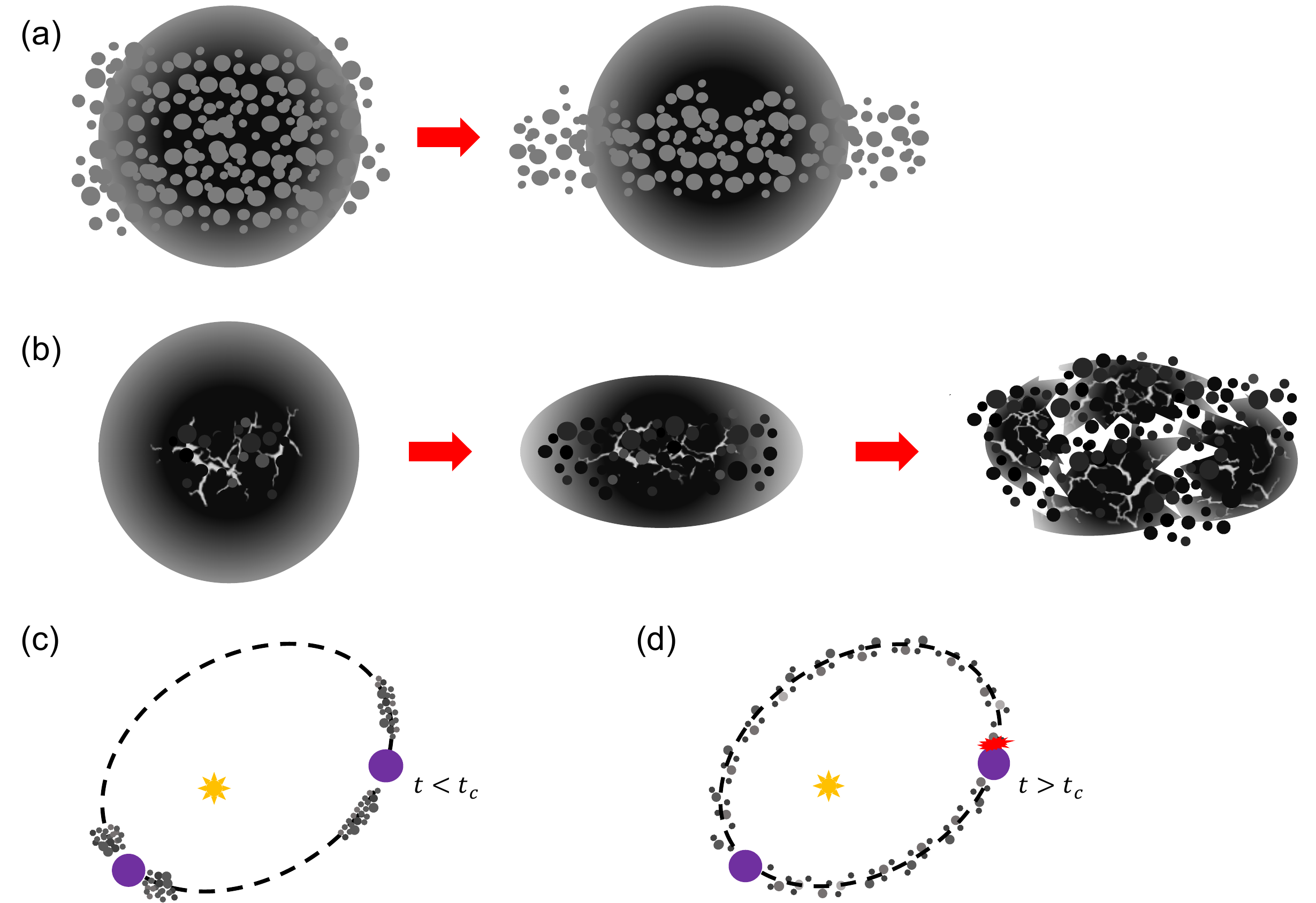}
    \caption{Schematic of Gault's structure and ejected dust particle evolution: (a) Surface structure failure, (b) Internal structure failure, (c) Orbital evolution of dust particles not perturbed by SRP and PRD ($Q_{pr}$ = 0), and (d) Orbital evolution of dust particles perturbed by SRP and PRD ($Q_{pr}$ $>$ 0). For the $Q_{pr}$ = 0 case, none of the ejected particles came back on the time scales we tested; i.e. the times we considered were less than its critical time. For $Q_{pr}$ $>$ 0, the particles are fully distributed along the Gault's orbit after $t_c$ ranging between 700 - 5,300 years (See the online animated figure included in the Appendix). Here, $t_c$ strongly depends on $Q_{pr}$.
    }
    \label{Fig4: evolution}
\end{figure}

\appendix
\section{Critical cohesive strength distribution for non-spherical cases}
\label{App:non_spherical}

\begin{figure}[h!]
    \centering
    \includegraphics[width=0.8\linewidth]{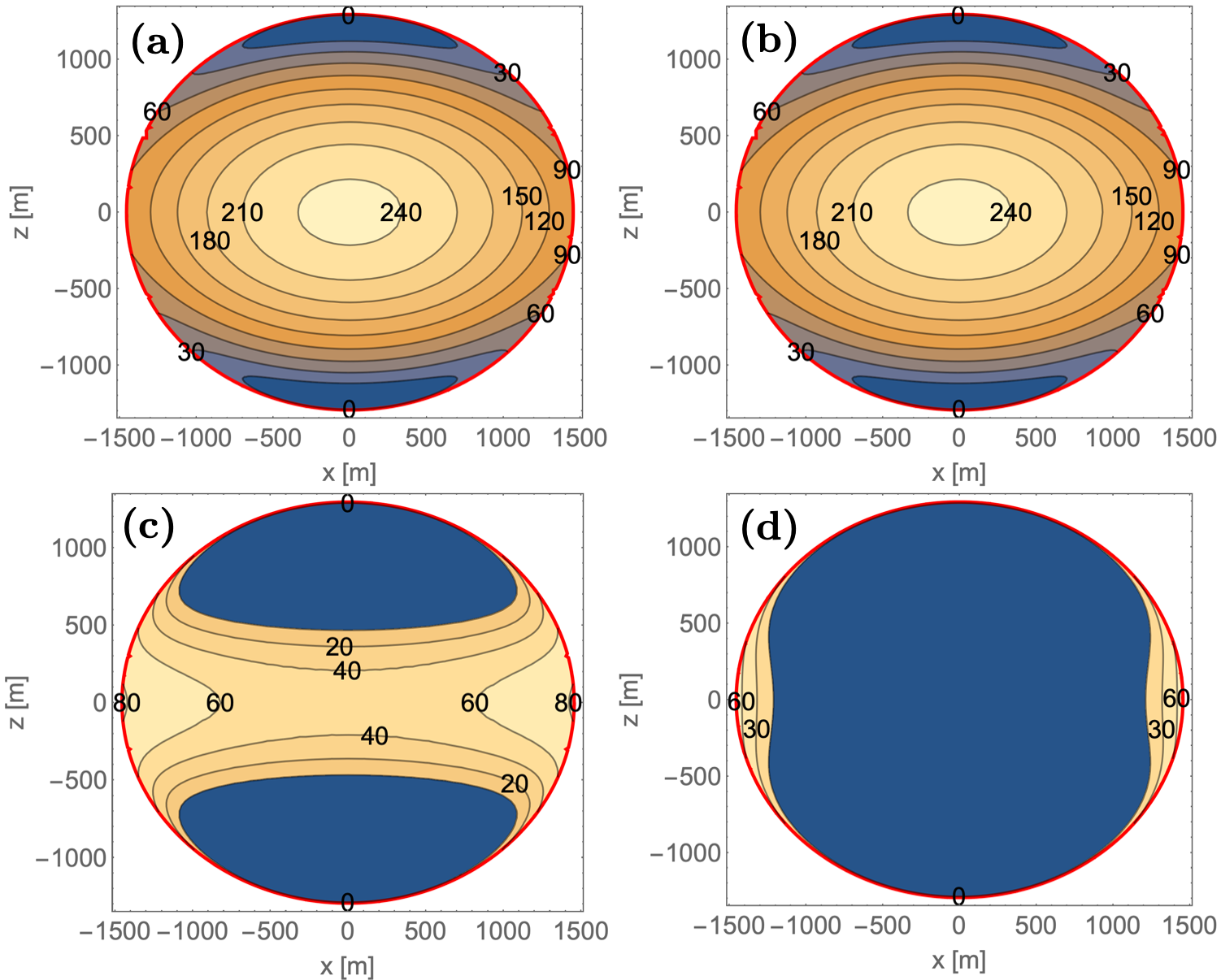}
    \caption{Distribution of $Y^*$ at the $x$-$z$ plane. The oblateness is 0.889, which is the same as Bennu’s. The spin period is $P = 2.5$ hr. Panels a, b, c, and d describe the bulk densities of 1.50, 1.75, 2.50, and 3.00 \textrm{g\ {cm}\textsuperscript{-3}}.}
    \label{fig5:Supplemental1}
\end{figure}

\begin{figure}
    \centering
    \includegraphics[width=0.8\linewidth]{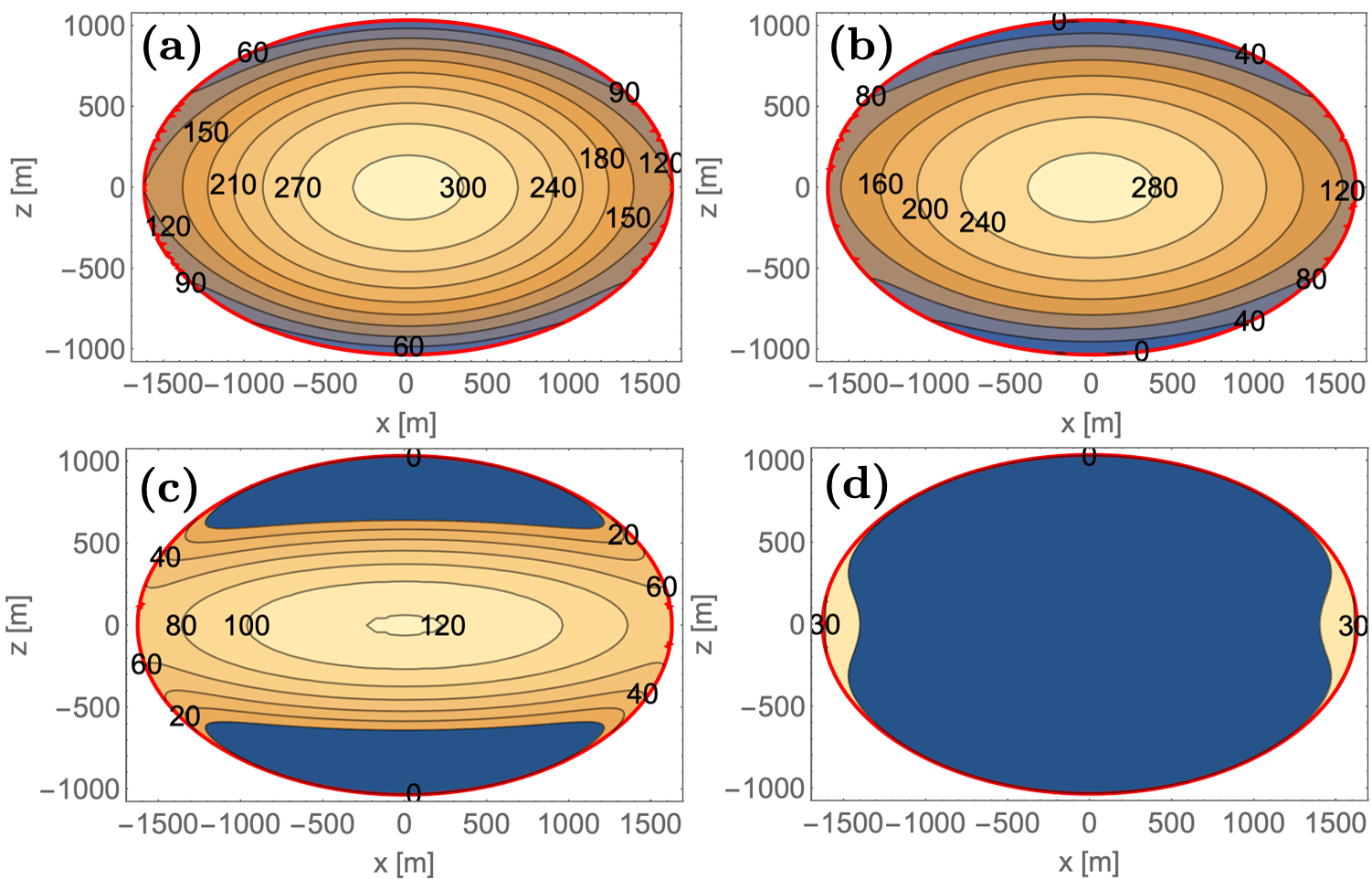}
    \caption{Distribution of $Y^*$ at the $x$-$z$ plane. The oblateness is 0.635, which is the same as Betulia’s.  The spin period is $P = 2.5$ hr. Panels a, b, c, and d describe the bulk densities of 1.50, 1.75, 2.50, and 3.00 \textrm{g\ {cm}\textsuperscript{-3}}.}
    \label{fig6:Supplemental2}
\end{figure}

The shape of Gault is not well constrained at present. In the main discussions, we assume the shape to be a sphere and investigate the critical cohesive strength distribution $Y^*$ at the current spin period by using the semi-analytical model (Figure \ref{Fig: structure}). However, Gault may have an elongated shape. Here, we test how $Y^*$ varies for such non-spherical bodies. We consider two sample cases of the oblateness, the ratio of the semi-minor axis to the semi-major axis: 0.889 for Bennu and 0.635 for asteroid (1580) Betulia \citep{hirabayashi2019rotationally}. The volume of Gault for these cases is kept the same with the sphere case. We do not account for the semi-intermediate axis for the oblateness to simplify the discussion. The same parameters listed in Table \ref{table:1} are used. Figure \ref{fig5:Supplemental1} shows the distribution of $Y^*$ on the $x$-$z$ plane for the 0.889 oblateness case. Comparing to the sphere case (Figure \ref{Fig: structure}), we observe $\sim 30$ Pa higher $Y^*$ values in the central region. The surface region shows the same or slightly lower $Y^*$, as the surface slope is milder than the sphere case. The 0.635 oblateness case (Figure \ref{fig6:Supplemental2}) shows a similar trend but with even higher $Y^*$ in the central region. We find that, in this range of oblateness, $\sim 40\%$ difference compared to the sphere case could arise. However, the failure modes inferred from Figures \ref{Fig: structure}, \ref{fig5:Supplemental1} and \ref{fig6:Supplemental2} are consistent and do not affect our conclusion. 

\section{Particle cloud orbital evolution}
\label{App:Particle_Ejection}
We present an animated figure of the time evolution of the 108 ejected particles from Gault for the epoch December 24, 2018 with a $Q_{pr}$ value of 2.

\begin{figure}[h!]
    \centering
    \includegraphics[width=0.45\linewidth]{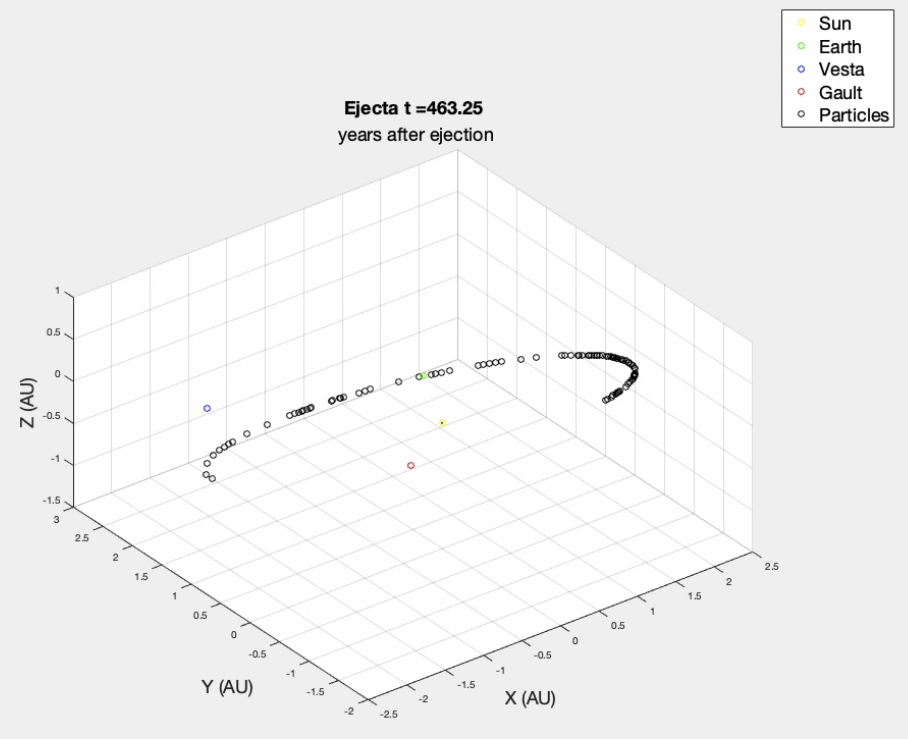}
    \caption{Animation of the orbital evolution of 108 particles ejected radially out from Gault at the epoch December 24, 2018 at velocity 0.15 m s$^{-1}$ and $Q_{pr}$ = 2. Prograde motion is counter-clockwise and retrograde motion is clockwise. At the beginning of the animation, the particles are ejected in all directions away from Gault, though they all get quickly blown behind (or retrograde) relative to Gault (though their heliocentric motion is prograde). Over the course of the animation, they slowly spread out and populate Gault's orbit and eventually return to Gault at a critical time of 542 years. See Figures \ref{Fig3: particle_dist}(c) and \ref{Fig4: evolution}(d).}
    \label{fig7:animation}
\end{figure}

\newpage
\bibliography{references}
\bibliographystyle{aasjournal}

\end{document}